\documentclass[12pt]{article}

\usepackage{amsfonts}
\usepackage{amsmath}
\usepackage{amsthm}
\usepackage{cite}
\usepackage{epsfig}
\usepackage{latexsym}
\usepackage{paralist}
\usepackage{fancyhdr}
\usepackage{graphicx}
\numberwithin{equation}{section}
\usepackage[vcentermath]{youngtab}
\usepackage{young}
\usepackage{ytableau}
\usepackage{etex}
\usepackage{braket}
\usepackage{float}
\usepackage{autobreak}

\usepackage{pict2e}   
\usepackage{animate}  
\usepackage{multirow}
\usepackage{bigdelim}
\usepackage{fancybox}
\usepackage{cals}

\def\be{\begin{equation}}
\def\ee{\end{equation}}
\def\bea{\begin{eqnarray}}
\def\eea{\end{eqnarray}}

\renewcommand{\thefootnote}{\fnsymbol{footnote}}

\begin{document}

\hfuzz=100pt
\title{{\Large \bf{3d s-confinement for three-index matters } }}
\date{}
\author{ Keita Nii$^a$\footnote{nii@itp.unibe.ch}
%, 
%Yuta Sekiguchi$^a$\footnote{yuta@itp.unibe.ch}
% and others$^{c,d}$
}
\date{\today}

\maketitle

\thispagestyle{fancy}
%\rhead{****-**-**}
\cfoot{}
\renewcommand{\headrulewidth}{0.0pt}

\vspace*{-1cm}
\begin{center}
%  \spa{0.5} \\
$^{a}${{\it Albert Einstein Center for Fundamental Physics }}
\\{{\it Institute for Theoretical Physics
}}
\\ {{\it University of Bern}}  
\\{{\it  Sidlerstrasse 5, CH-3012 Bern, Switzerland}}
%\\ {{\it  }}
 % \spa{0.5} \\
%$^b${{\it Department of Physics}}
%\\ {{\it Nagoya University, Nagoya 464-8602, Japan}}
%\spa{0.5}  \\
%$^c${{\it  }}
%\\ {{\it }}
%\spa{0.5}  \\
%$^d${{\it }}

\end{center}

\begin{abstract}
We present s-confinement phases for three-index matters in three-dimensional supersymmetric gauge theories. We find that the 3d $\mathcal{N}=2$ $SU(6)$ and $USp(6)$ gauge theories with three-index anti-symmetric matters show confining phases. The exact superpotentials which describe their low-energy dynamics are derived. We check the validity of our analysis in various ways, including superconformal indices and some deformations. 
\end{abstract}

\renewcommand{\thefootnote}{\arabic{footnote}}
\setcounter{footnote}{0}

\newpage
%\tableofcontents 
%\clearpage

%%%%%%%%%%%%%%%%%%%%%%%%%%%%%%%%%%%%%%%%%%%%%%%%%%%%%%%%%%%
%%%%%%%%%%%%%%%%%%%%%%%%%%%%%%%%%%%%%%%%%%%%%%%%%%%%%%%%%%%
%%%%%%%%%%%%%%%%%%%%%%%%%%%%%%%%%%%%%%%%%%%%%%%%%%%%%%%%%%%
%%%%%%%%%%%%%%%%%%%%%%%%%%%%%%%%%%%%%%%%%%%%%%%%%%%%%%%%%%%
\section{Introduction}
%%%%%%%%%%%%%%%%%%%%%%%%%%%%%%%%%%%%%%%%%%%%%%%%%%%%%%%%%%%
%%%%%%%%%%%%%%%%%%%%%%%%%%%%%%%%%%%%%%%%%%%%%%%%%%%%%%%%%%%
%%%%%%%%%%%%%%%%%%%%%%%%%%%%%%%%%%%%%%%%%%%%%%%%%%%%%%%%%%%
%%%%%%%%%%%%%%%%%%%%%%%%%%%%%%%%%%%%%%%%%%%%%%%%%%%%%%%%%%%

%[general Intro, matter contentsによってphaseが変わる, ]
In supersymmetric gauge theories, one can exactly study the low-energy dynamics by employing power of holomorphy and various non-renoramlaization theorems \cite{Seiberg:1994bz,Seiberg:1994pq,Seiberg:1997vw}. The perturbative corrections are severely controlled and non-perturbative corrections from instantons are derived in a reliable way. We are interested in strongly-coupled theories whose low-energy limit generally allows various phases depending on the matter contents. One of the most fascinating phases is a confinement phase. Supersymmetric gauge theories sometimes exhibit so-called s-confinement which is a confining phase without (global) symmetry breaking. In four spacetime dimension, various s-confinement phases are constructed for fundamental, anti-symmetric and three-index anti-symmetric matters (see \cite{Csaki:1996zb} for classical groups). In three spacetime dimension, the s-confinement is found for fundamental and anti-symmetric matters (see, for example, \cite{Aharony:1997bx,Csaki:2014cwa,Amariti:2015kha,Nii:2016jzi}). In this paper, we study the s-confinement phases for the 3d SUSY gauge theories with three-index matters.

%[4d, multi-index matter, 2-index matter]
It is generally difficult to study the low-energy dynamics of the SUSY gauge theory with multi-index matters such as adjoint matters, (anti-)symmetric tensors or matters with more involved young tableaus. There are two ways to study such theories. One way is to introduce a superpotential for multi-index matters, which truncates the chiral ring and simplifies the dynamics. The other way is to use a de-confinement technique \cite{Berkooz:1995km,Pouliot:1995me,Luty:1996cg,Nii:2016jzi}. In this technique, we can think of the multi-index matters as mesons or baryons of some confining gauge theories. Hence, in the UV region, we obtain a product gauge group theory with (bi-)fundamental matters, which is more tractable than the original theory with multi-index matters. For two-index matters, the de-confinement technique is very effective, but for matters with more than two indices it is not available.

%[In this paper]
We will tackle with a problem of constructing the s-confinement phases for the theory with multi-index matters, especially three-index anti-symmetric matters. In four spacetime dimension, the s-confinement for the three-index matters are restricted to the $SU(6)$ gauge theory with a single third-order antisymmetric tensor and four (anti-)fundamental flavors \cite{Csaki:1996zb,Csaki:1997cu}. We will search for the three-dimensional s-confinement for three-index matters. We will find that the 3d s-confinement phases for three-index matters are more richer than 4d. In this paper, we consider two theories: One is a 3d $\mathcal{N}=2$ $SU(6)$ gauge theory with a single third-order antisymmetric tensor and three (anti-)fundamentals. The theory is similar to the 4d one and actually the 3d s-confinement phase is obtained from the 4d description via a real mass deformation. The other is a 3d $\mathcal{N}=2$ $USp(6)$ gauge theory with a third-order antisymmetric tensor and three fundamentals, which has no 4d counterpart.

%[Organization]
 The rest of this paper is organized as follows. 
 In Section 2, we discuss the 3d $\mathcal{N}=2$ $SU(6)$ gauge theory with a three-index matter. The corresponding 4d theory is also reviewed. We consider the relation between the 3d and 4d theories. 
In Section 3, we move on to the s-confinement phase for the 3d $\mathcal{N}=2$ $USp(6)$ gauge theory with a three-index matter. We will compute the superconformal index as a consistency check of our analysis.
In Section 4, we will summarize the results and discuss possible future directions to be studied.

\newpage
%%%%%%%%%%%%%%%%%%%%%%%%%%%%%%%%%%%%%%%%%%%%%%%%%%%%%%%%%%%
%%%%%%%%%%%%%%%%%%%%%%%%%%%%%%%%%%%%%%%%%%%%%%%%%%%%%%%%%%%
%%%%%%%%%%%%%%%%%%%%%%%%%%%%%%%%%%%%%%%%%%%%%%%%%%%%%%%%%%%
%%%%%%%%%%%%%%%%%%%%%%%%%%%%%%%%%%%%%%%%%%%%%%%%%%%%%%%%%%%
\section{$SU(6)$ gauge theories with three-index matters}
%%%%%%%%%%%%%%%%%%%%%%%%%%%%%%%%%%%%%%%%%%%%%%%%%%%%%%%%%%%
%%%%%%%%%%%%%%%%%%%%%%%%%%%%%%%%%%%%%%%%%%%%%%%%%%%%%%%%%%%
%%%%%%%%%%%%%%%%%%%%%%%%%%%%%%%%%%%%%%%%%%%%%%%%%%%%%%%%%%%
%%%%%%%%%%%%%%%%%%%%%%%%%%%%%%%%%%%%%%%%%%%%%%%%%%%%%%%%%%%
In this section, we will discuss the s-confinement phases for the 3d and 4d supersymmetric $SU(6)$ gauge theories with a three-index anti-symmetric matter. Since the 4d s-confinement for three-index matters was already constructed in \cite{Csaki:1996zb} for the $SU(6)$ case, we first briefly review it. By dimensionally reducing the 4d theory onto 3d via circle compactification, the 4d theory leads to the 3d s-confinement. We will directly analyze the 3d theory in the next subsection.

%%%%%%%%%%%%%%%%%%%%%%%%%%%%%%%%%%%%%%%%%%%%%%%%%%%%%%%%%%%
%%%%%%%%%%%%%%%%%%%%%%%%%%%%%%%%%%%%%%%%%%%%%%%%%%%%%%%%%%%
\subsection{4d $\mathcal{N}=1$ $SU(6)$ with a three-index matter}
%%%%%%%%%%%%%%%%%%%%%%%%%%%%%%%%%%%%%%%%%%%%%%%%%%%%%%%%%%%
%%%%%%%%%%%%%%%%%%%%%%%%%%%%%%%%%%%%%%%%%%%%%%%%%%%%%%%%%%%
We first consider the 4d $\mathcal{N}=1$ $SU(6)$ gauge theory with a three-index anti-symmetric matter and four (anti-)fundamental flavors \cite{Csaki:1996zb,Csaki:1997cu}. The theory is known to be s-confining in a far-infrared limit. Table \ref{T4dSU6} shows the matter contents and their quantum numbers. The global symmetries are $SU(4)_L \times SU(4)_R \times U(1)_B \times U(1)_A \times U(1)' \times U(1)_R$ whose subgroup, $U(1)_A \times U(1)' \times U(1)_R$ part, is anomalous due to the chiral anomalies in 4d. Therefore, the dynamical scale $\eta =\Lambda^b$ is also charged under these $U(1)$ symmetries, where $b$ is a one-loop beta function coefficient. Since we are eventually interested in a corresponding 3d theory, we will use these spurious symmetries in what follows. 

\begin{table}[H]\caption{Quantum numbers of 4d $\mathcal{N}=1$ $SU(6)$ with ${\tiny \protect\yng(1,1,1)}$ and $4\,({\tiny \protect\yng(1)} +{\tiny \overline{\protect\yng(1)}})$} 
\begin{center}
\scalebox{1}{
  \begin{tabular}{|c||c||c|c|c|c|c|c| } \hline
  &$SU(6)$&$SU(4)_L$&$SU(4)_R$&$U(1)_B$ & $U(1)_A$ & $U(1)'$ &$U(1)_R$ \\ \hline
$Q$& ${\tiny \yng(1)}$ & ${\tiny \yng(1)}$&1&$1$&1&  0  &$R_Q$ \\
$\tilde{Q}$& $\bar{{\tiny \yng(1)}}$ &1& ${\tiny \yng(1)}$&$-1$&1& 0 &$R_Q$ \\
$A$& ${\tiny \yng(1,1,1)}$ &1&1&0&$0$&$1$ & $R_A$ \\[6pt] \hline
$\eta:=\Lambda^b$ &1&1&1&0&8&6 &$8R_Q +6R_A-2$\\ \hline 
$M_0:=Q \tilde{Q}$&1&${\tiny \yng(1)}$&${\tiny \yng(1)}$&0&2&0&$2R_Q$ \\
$M_2:=QA^2 \tilde{Q}$&1&${\tiny \yng(1)}$&${\tiny \yng(1)}$&0&$2$&2 &$2R_Q+2R_A$\\
$B_1:=AQ^3$&1&$\bar{{\tiny \yng(1)}}$&1&3&3&1 & $3R_Q+R_A$ \\
$\bar{B}_1:=A \tilde{Q}^3$&1&1&$\bar{{\tiny \yng(1)}}$&$-3$&3&1 &$3R_Q+R_A$  \\
$B_3:=A^3Q^3$ &1&$\bar{{\tiny \yng(1)}}$&1&3&$3$&3 &$3R_Q+3R_A$\\
$\bar{B}_3:=A^3 \tilde{Q}^3$ &1&1&$\bar{{\tiny \yng(1)}}$&$-3$&$3$&0 & $3R_Q+3R_A$\\
$T:=A^4$&1&1&1&0&$0$&$4$&$4R_A$ \\ \hline
  \end{tabular}}
  \end{center}\label{T4dSU6}
\end{table}

In order to describe the Higgs branch of the moduli space of vacua, we introduced the following composite operators
\begin{gather}
M_0:=Q \tilde{Q},~~~M_2:=QA^2 \tilde{Q},~~~T:=A^4, \nonumber \\
B_1:=AQ^3,~~~\bar{B}_1:=A \tilde{Q}^3,~~~B_3:=A^3Q^3,~~~\bar{B}_3:=A^3 \tilde{Q}^3. \label{SU6Higgs}
\end{gather}
These variables are not independent of each other and they are constrained. These (classical) constraints are depicted from the following superpotential
\begin{align}
W= \frac{1}{\eta} \left( M_0 B_1 \bar{B}_1 T+B_3 \bar{B}_3 M_0 +M_2^3 M_0 +T M_2M_0^3+\bar{B}_1B_3M_2+B_1\bar{B}_3 M_2  \right), \label{4dSU6}
\end{align}
where $\eta$ is inserted to have the correct charges of the superpotential and we omitted the relative coefficients for simplicity.

Since the dual description \eqref{4dSU6} has no gauge interaction, it is quite simple to dimensionally reduce the theory to 3d \cite{Aharony:2013dha,Csaki:2014cwa}. By putting the theory on a circle and taking a small circle limit, the theory flows to the 3d s-confined phase. In order to obtain the theory without monopole superpotential on the electric side, we have to introduce the real masses by background-gauging the $SU(4)_L\times SU(4)_R \times U(1)_B $ symmetries and by giving the expectation values to the scalar modes of the background vector superfields. In this deformation, the monopole superpotential on the electric side drops off and we obtain the 3d $\mathcal{N}=2$ $SU(6)$ gauge theory with ${\tiny \yng(1,1,1)}$ and $3~({\tiny \yng(1)} +{\tiny \overline{\yng(1)}})$ without superpotential. For fundamental chiral multiplets, we introduce the real masses as follows. 
\begin{align}
&\begin{pmatrix}
0 & && \\
 & 0& &\\
 &&0& \\
 &&&m 
\end{pmatrix} = \frac{mQ_B}{4}I+\begin{pmatrix}
\frac{m}{4} & && \\
 & \frac{-m}{4}& &\\
 &&0& \\
 &&&0 
\end{pmatrix}+\begin{pmatrix}
\frac{m}{4} & && \\
 & 0& &\\
 &&\frac{-m}{4}& \\
 &&&0 
\end{pmatrix}-\begin{pmatrix}
\frac{3m}{4} & && \\
 & 0& &\\
 &&0& \\
 &&&\frac{3m}{4}
\end{pmatrix} 
\end{align}
where these matrices act on the flavor indices. $Q_B$ is a global $U(1)_B$ charge of the fundamental multiplet and it is $Q_B=1$ for fundamental representations. Hence, the last flavor of the (anti-)fundamental multiplets is integrated out. On the dual side, this deformation gives the real masses for the confined chiral superfields. We have to keep the following fields in the low-energy limit 
\begin{align}
M_0=&\begin{pmatrix}
 & && 0\\
 & M^{3d}_0& &0\\
 &&& 0 \\
 0&0&0& Y 
\end{pmatrix},~~M_2=\begin{pmatrix}
 & && 0\\
 & M^{3d}_2& &0\\
 &&& 0 \\
 0&0&0& \tilde{Y} 
\end{pmatrix} \\
B_1&=\begin{pmatrix}
 0& 0&0& B_1^{3d}
 \end{pmatrix},~~\bar{B}_1=\begin{pmatrix}
 0& 0&0& \bar{B}_1^{3d} 
 \end{pmatrix},\\
 B_3&=\begin{pmatrix}
 0& 0&0& B_3^{3d}
 \end{pmatrix},~~~\bar{B}_3=\begin{pmatrix}
 0& 0&0& \bar{B}_3^{3d}
 \end{pmatrix},
\end{align}
where we have renamed the bottom components of the mesonic fields into $Y$ and $\tilde{Y}$ because these will be identified with the Coulomb branch operators in 3d.
In this redefinition, the superpotential reduces to 
\begin{align}
W=Y \left( \det \, M_2 +TM_0^2M_2 +TB_1 \bar{B}_1 +B_3 \bar{B}_3 \right) +\tilde{Y} \left(T \det \, M_0 +M_0 M_2^2+B_1 \bar{B}_3 +\bar{B}_1 B_3  \right), \label{4d3d}
\end{align}
where we omitted the 3d labels and absorbed the dynamcal scale into the fields for simplicity. The s-confined description \eqref{4d3d} is equivalent to the 3d $\mathcal{N}=2$ $SU(6)$ gauge theory with ${\tiny \yng(1,1,1)}$ and $3~({\tiny \yng(1)} +{\tiny \overline{\yng(1)}})$.
We will reproduce this superpotential in the next subsection by directly analyzing the 3d theory.

Before moving on to the 3d story, let us consider the 4d $\mathcal{N}=1$ $SU(6)$ gauge theory with a three-index anti-symmetric matter and three (anti-)fundamental flavors, which was also studied in \cite{Csaki:1996zb}. The theory can be obtained via the complex mass deformation for a (anti-)fundamental matter. Table \ref{4dSU63flavor} shows the matter contents, the moduli coordinates and their quantum numbers.

\begin{table}[H]\caption{Quantum numbers of 4d $\mathcal{N}=1$ $SU(6)$ with ${\tiny \protect\yng(1,1,1)}$ and $3\,({\tiny \protect\yng(1)} +{\tiny \overline{\protect\yng(1)}})$} 
\begin{center}
\scalebox{1}{
  \begin{tabular}{|c||c||c|c|c|c|c|c| } \hline
  &$SU(6)$&$SU(3)_L$&$SU(3)_R$&$U(1)_B$ & $U(1)_A$ & $U(1)'$ &$U(1)_R$ \\ \hline
$Q$& ${\tiny \yng(1)}$ & ${\tiny \yng(1)}$&1&$1$&1&  0  &$R_Q$ \\
$\tilde{Q}$& $\bar{{\tiny \yng(1)}}$ &1& ${\tiny \yng(1)}$&$-1$&1& 0 &$R_Q$ \\
$A$& ${\tiny \yng(1,1,1)}$ &1&1&0&$0$&$1$ & $R_A$ \\[6pt] \hline
$\eta:=\Lambda^b$ &1&1&1&0&6&6 &$6R_Q +6R_A$\\ \hline 
$M_0:=Q \tilde{Q}$&1&${\tiny \yng(1)}$&${\tiny \yng(1)}$&0&2&0&$2R_Q$ \\
$M_2:=QA^2 \tilde{Q}$&1&${\tiny \yng(1)}$&${\tiny \yng(1)}$&0&$2$&2 &$2R_Q+2R_A$\\
$B_1:=AQ^3$&1&1&1&3&3&1 & $3R_Q+R_A$ \\
$\bar{B}_1:=A \tilde{Q}^3$&1&1&1&$-3$&3&1 &$3R_Q+R_A$  \\
$B_3:=A^3Q^3$ &1&1&1&3&$3$&3 &$3R_Q+3R_A$\\
$\bar{B}_3:=A^3 \tilde{Q}^3$ &1&1&1&$-3$&$3$&0 & $3R_Q+3R_A$\\
$T:=A^4$&1&1&1&0&$0$&$4$&$4R_A$ \\ \hline
  \end{tabular}}
  \end{center}\label{4dSU63flavor}
\end{table}

In this case, the Higgs branch operators need two constraints and one of them is quantum-mechanically modified. The constrains are realized by the Lagrange multipliers $X_{1,2}$ as 
\begin{align}
W=X_1 \left( B_1 \bar{B}_1 T+B_3 \bar{B}_3 +M_2^3 +TM_2M_0^2 +\eta \right) +X_2 \left( M_2^2M_0 +TM_0^3 +\bar{B}_1B_3+B_1 \bar{B}_3\right). \label{4dQC}
\end{align}
Notice the resemblance between \eqref{4d3d} and \eqref{4dQC}. 
The equation of motion for $X_1$ leads to the symmetry breaking of the global symmetry. We will reproduce this result from the 3d theory point of view in a next subsection.

%%%%%%%%%%%%%%%%%%%%%%%%%%%%%%%%%%%%%%%%%%%%%%%%%%%%%%%%%%
%%%%%%%%%%%%%%%%%%%%%%%%%%%%%%%%%%%%%%%%%%%%%%%%%%%%%%%%%%
\subsection{3d $\mathcal{N}=2$ $SU(6)$ with a single three-index matter}
%%%%%%%%%%%%%%%%%%%%%%%%%%%%%%%%%%%%%%%%%%%%%%%%%%%%%%%%%%
%%%%%%%%%%%%%%%%%%%%%%%%%%%%%%%%%%%%%%%%%%%%%%%%%%%%%%%%%%
Let us move on to the analysis of the 3d $\mathcal{N}=2$ $SU(6)$ gauge theory with a three-index anti-symmetric matter and three (anti-)fundamental flavors, whose Lagrangian is just obtained via the dimensional reduction of the 4d theory discussed in a previous subsection. The global symmetries are identical to the 4d ones but all the $U(1)$ symmetries are not anomalous. Table \ref{3dSU6} shows the matter contents and their global charges. The Higgs branch is described by the same composite operators as \eqref{SU6Higgs}.

\begin{table}[H]\caption{Quantum numbers of 3d $\mathcal{N}=2$ $SU(6)$ with a three-index matter} 
\begin{center}
\scalebox{0.91}{
  \begin{tabular}{|c||c||c|c|c|c|c|c| } \hline
 & $SU(6)$ & $SU(3)$ & $SU(3)$ & $U(1)_B$ & $U(1)_A$ & $U(1)'$ & $U(1)_R$ \\  \hline
 $Q$ & ${\tiny \yng(1)} $ &${\tiny \yng(1)} $&1&1&1&0&$R_Q$ \\
 $\tilde{Q}$& ${\tiny \bar{\yng(1)}} $ &1&${\tiny \bar{\yng(1)}} $&$-1$&1&0&$R_Q$ \\
 $A$&${\tiny \yng(1,1,1)} $&1&1&0&0&1&$R_A$  \\[6pt] \hline
 $Y$&1&1&1&0&$-6$&$-6$& $-10-6 (R_Q-1)-6(R_A-1) $ \\
 $\tilde{Y}$ &1&1&1&0&$-6$&$-4$&$-8-6(R_Q-1) -4(R_A-1) $ \\ \hline
 %$\hat{Y}$&1&1&1&0&$-2N_f$&$-4$& $ -6- 2N_f(R_Q-1) -4 (R_A-1)$ \\ \hline 
$M_0:=Q\tilde{Q}$ &1&${\tiny \yng(1)} $& ${\tiny \bar{\yng(1)}} $&0&2&0&$2R_Q$ \\
$M_2:=QA^2\tilde{Q}$ &1&${\tiny \yng(1)} $& ${\tiny \bar{\yng(1)}} $&0&2&2&$2R_Q+2R_A$ \\
 $B_1:=AQ^3$&1&$1 $&1&3&3&1& $3R_Q+R_A$\\
$\bar{B}_1:=A\tilde{Q}^3$ &1&1&$1$&$-3$&3&1& $3R_Q+R_A$ \\
$B_3:=A^3Q^3$ &1&$1 $&1&3&3&3& $3R_Q+3R_A$ \\
 $\bar{B}_3:=A^3 \tilde{Q}^3$&1&1&$1 $&$-3$&3&3& $3R_Q+3R_A$ \\
 $T:=A^4$ &1&1&1&0&0&$4$&$4R_A$ \\ \hline
 %$\det \, M_0$&1&1&1&0&$2N_f$&0&$2N_fR_Q$ \\
 %$\det \, M_2$ &1&1&1&0&$2N_f$&$2N_f$&$2N_f(R_Q+R_A)$ \\ \hline
  \end{tabular}}
  \end{center}\label{3dSU6}
\end{table}

The classical Coulomb branch is parametrized by the following coordinates
\begin{align}
Y_i \simeq \exp (\sigma_i -\sigma_{i+1})~~~~(i=1,\cdots,5),
\end{align}
where $\sigma_i$ are the diagonal components of the adjoint scalar in the $SU(6)$ vector superfield. We omitted the dependence of the gauge coupling and the dual photons for simplicity. These (classical) flat directions are generally lifted by non-perturbative effects from the monopoles and some directions remain flat at a quantum level. The Coulomb branch becomes multi-dimensional since the theory contains the multi-index matters \cite{Csaki:2014cwa,Amariti:2015kha}. Therefore, various combinations of the classical coordinates should be studied separately. 

The first candidate of the quantum Coulomb moduli is $Y= \prod_{i=1}^5 Y_i$, whose vev induces the higgsing $SU(6) \rightarrow SU(4) \times U(1)_1 \times U(1)_2$. The matter fields are decomposed as
\begin{align}
{\tiny \yng(1)} & \rightarrow  {\tiny \yng(1)}_{(0,-1)}+ \mathbf{1}_{(1,2)}+\mathbf{1}_{(-1,2)} ~~~~&( \mathbf{6} \rightarrow \mathbf{4}+ \mathbf{1}+ \mathbf{1}), \\
{\tiny \bar{\yng(1)}} & \rightarrow {\tiny \bar{\yng(1)}}_{(0,1)}+ \mathbf{1}_{(-1,-2)}+\mathbf{1}_{(1,-2)} ~~~~&( \bar{\mathbf{6}} \rightarrow \bar{\mathbf{4}}+ \mathbf{1}+ \mathbf{1}), \\
{\tiny \yng(1,1,1)} & \rightarrow  {\tiny \yng(1,1)}_{(1,0)}+{\tiny \yng(1,1)}_{(-1,0)}+ {\tiny \bar{\yng(1)}}_{(0,-3)}+{\tiny \yng(1)}_{(0,3)}   &(\mathbf{20} \rightarrow \mathbf{6}+ \mathbf{6}+ \bar{\mathbf{4}}+ \mathbf{4}).
\end{align}
From this decomposition, we can compute the effective Chern-Simons levels between $U(1)_1$ and other global $U(1)$ symmetries
\begin{align}
k_{eff}^{U(1)_1 U(1)_{global}} =3 Q_{\Yboxdim{6pt} \yng(1)}+3 Q_{\Yboxdim{6pt} \overline{\yng(1)}} +6 Q_{\tiny \Yboxdim{6pt} \yng(1,1,1)} + 10 Q_{adj}.
\end{align}
The quantum numbers of $Y$ can be computed from this mixed CS terms (see Table \ref{3dSU6}). The $Y$ coordinate is globally defined when the theory only contains the (anti-)fundamental flavors \cite{Aharony:1997bx}. Hence, we assume that $Y$ is one of the Coulomb branch coordinates. 

The another candidate is $\tilde{Y} :=\sqrt{Y_1Y_2^2Y_3^2Y_4^2Y_5}$ as in \cite{Amariti:2015kha}, which induces the gauge symmetry breaking $SU(6) \rightarrow SU(2)_t \times SU(2)_m \times SU(2)_b \times U(1)'_1 \times U(1)'_2$. The matter fields are decomposed as
\begin{align}
{\tiny \yng(1)} & \rightarrow ({\tiny \yng(1)},\cdot, \cdot)_{(1,1)} +(\cdot , {\tiny \yng(1)} , \cdot)_{(0,-2)} +(\cdot,\cdot,{\tiny \yng(1)})_{(-1,1)}, \\
{\tiny \bar{\yng(1)}} & \rightarrow  ({\tiny \yng(1)},\cdot, \cdot)_{(-1,-1)} +(\cdot , {\tiny \yng(1)} , \cdot)_{(0,2)} +(\cdot,\cdot,{\tiny \yng(1)})_{(1,-1)},\\
{\tiny \yng(1,1,1)} & \rightarrow ({\tiny \yng(1)},{\tiny \yng(1)},{\tiny \yng(1)})_{(0,0,0)} +({\tiny \yng(1)},\cdot,\cdot)_{(1,-3)} +({\tiny \yng(1)},\cdot,\cdot)_{(-1,3)}  \nonumber \\&\qquad+(\cdot,{\tiny \yng(1)},\cdot)_{(2,0)}+(\cdot,{\tiny \yng(1)},\cdot)_{(-2,0)} +(\cdot,\cdot,{\tiny \yng(1)})_{(-1,-3)}+(\cdot,\cdot,{\tiny \yng(1)})_{(1,3)}.
\end{align}
The mixed Chern-Simons term becomes
\begin{align}
k_{eff}^{U(1)_1 U(1)_{global}} = 3 Q_{\Yboxdim{6pt} \yng(1)}+ 3 Q_{\Yboxdim{6pt} \bar{\yng(1)}} +4 Q_{\tiny \Yboxdim{6pt} \yng(1,1,1)} +8Q_{adj}.
\end{align} 
From this expression, we can compute the quantum numbers of $\tilde{Y}$. We assume that these two operators, $Y$ and $\tilde{Y}$ are the correct coordinates for the quantum Coulomb moduli.

 Now, we listed all the moduli coordinates. One can immediately write down the superpotential consistent with all the symmetries in Table \ref{3dSU6}.
\begin{align}
W=Y \left( \det \, M_2 +TM_0^2M_2 +TB_1 \bar{B}_1 +B_3 \bar{B}_3 \right) +\tilde{Y} \left(T \det \, M_0 +M_0 M_2^2+B_1 \bar{B}_3 +\bar{B}_1 B_3  \right)
\end{align}
By introducing the superpotential from the KK monopole, $W= \eta Y$ and regarding the Coulomb branch coordinates as the Lagrange multipliers, one can reproduce the 4d quantum constraints \eqref{4dQC}. Furthermore the 3d superpotential is consistent with the previous result \eqref{4d3d} which was derived from 4d. This confirms our analysis of the Coulomb branch.

We can verify our above analysis by flowing to the various Higgs branch.
Let us consider the mesonic Higgs branch, where $\braket{M_0}$ gets non-zero expectation values. 
When $\braket{M_0}$ is rank-one, the theory flows to the 3d $\mathcal{N}=2$ $SU(5)$ gauge theory with an antisymmetric flavor and two fundamental flavors. Its low-energy dynamics is known to be s-confining \cite{Csaki:2014cwa,Nii:2016jzi}. We can alternatively turn on $M_2$ whose vev breaks the gauge group as $SU(6) \rightarrow USp(4)$. The UV theory leads to the 3d $\mathcal{N}=2$ $USp(4)$ gauge theory with one anti-symmetric and four fundamentals, which is again s-confining \cite{Nii:2016jzi}. In both cases, the low-energy theories along the Higgs branch are s-confining and have two-dimensional Coulomb branch. This is consistent with our analysis for the $SU(6)$ theory with a three-index matter.

\subsection{Superconformal Indices}
%%%%%%%%%%%%%%%%%%%%%%%%%%%%%%%%%%%%%%%%%%%%%%%%%%%%%%%%%%
%%%%%%%%%%%%%%%%%%%%%%%%%%%%%%%%%%%%%%%%%%%%%%%%%%%%%%%%%%

Since the $SU(6)$ gauge theory discussed in a previous subsection exhibits the s-confining phase, we can compare the superconformal indices for the electric (UV) and dual (IR) descriptions. This would be a non-trivial check of our analysis. For the precise definition of the superconformal indices, see \cite{Bhattacharya:2008bja,Kim:2009wb,Imamura:2011su,Imamura:2011uj,Kapustin:2011jm,Spiridonov:2009za,Bashkirov:2011vy,Kim:2013cma}
The dual index only has the contributions from the gauge singlets chiral superfields and takes the following form

\scriptsize 
\begin{align}
I_{dual} &=1+9 t^2 x^{1/4}+\sqrt{x} \left(\frac{1}{t^6 u^6}+45 t^4+2 t^3 u+9 t^2 u^2+u^4\right)+x^{3/4} \left(\frac{1}{t^6 u^4}+165 t^6+18 t^5 u+\frac{9}{t^4 u^6}+81 t^4 u^2+2 t^3 u^3+9 t^2 u^4\right) \nonumber \\
&\qquad +x \left(\frac{1}{t^{12} u^{12}}+495 t^8+90 t^7 u+408 t^6 u^2+\frac{1}{t^6 u^2}+36 t^5 u^3+90 t^4 u^4+\frac{18}{t^4 u^4}+2 t^3 u^5+\frac{2}{t^3 u^5}+9 t^2 u^6+\frac{45}{t^2 u^6}+u^8\right)\nonumber \\
&\qquad+x^{5/4} \biggl(\frac{1}{t^{12} u^{10}}+\frac{9}{t^{10} u^{12}}+1287 t^{10}+330 t^9 u+1512 t^8 u^2+252 t^7 u^3+573 t^6 u^4+\frac{1}{t^6}  \nonumber \\
&\qquad \qquad \qquad \qquad \qquad \qquad \qquad \qquad   +36 t^5 u^5+81 t^4 u^6+\frac{18}{t^4 u^2}+2 t^3 u^7 +\frac{2}{t^3 u^3}+9 t^2 u^8+\frac{126}{t^2 u^4}+\frac{18}{t u^5}+\frac{165}{u^6}\biggr) +\cdots, 
\end{align}
\normalsize

\noindent where $t$ and $u$ are the fugacities for the $U(1)_A$ and $U(1)'$ symmetries respectively. We set $R_Q=R_A =\frac{1}{8}$ for simplicity.

On the other hand, the electric index is decomposed into the indices with different GNO charges. Since the gauge group is $SU(6)$, the magnetic charges are parametrized by $(m_1,m_2,m_3,m_4,m_5,m_6)$ with a constraint $\sum_{i=1}^{6} m_i=0$. The lower-order indices are obtained as follows.

\begin{align}
I_{electric}^{(0,0,0,0,0,0)} &= 1+9 t^2 x^{1/4}+\sqrt{x} \left(45 t^4+2 t^3 u+9 t^2 u^2+u^4\right)+\cdots \\
I_{electric}^{\left(\frac{1}{2},0,0,0,0,\frac{-1}{2} \right)} &=\frac{\sqrt{x}}{t^6 u^6}+x^{3/4} \left(\frac{1}{t^6 u^4}+\frac{9}{t^4 u^6}\right)+\frac{x \left(45 t^4+2 t^3 u+18 t^2 u^2+u^4\right)}{t^6 u^6} \nonumber \\
&\qquad +x^{5/4} \left(\frac{1}{t^6}+\frac{18}{t^4 u^2}+\frac{2}{t^3 u^3}+\frac{126}{t^2 u^4}+\frac{18}{t u^5}+\frac{165}{u^6}\right)+\cdots \\
I_{electric}^{\left(1,0,0,0,0, -1 \right)} &=\frac{x}{t^{12} u^{12}}+\frac{x^{5/4} \left(9 t^2+u^2\right)}{t^{12} u^{12}}+\frac{x^{3/2} \left(45 t^4+2 t^3 u+18 t^2 u^2+u^4\right)}{t^{12} u^{12}} +\cdots
\end{align}

The sector with zero GNO charge includes only the Higgs branch operators. The second term $9 t^2 x^{1/4}$ is identified with the mesonic operator $M_0$ which has nine components. The third term $\sqrt{x} \left(45 t^4+2 t^3 u+9 t^2 u^2+u^4\right)$ consists of five operators; $M_0^2, B_1, \bar{B}_1, M_2$ and $T$, which is consistent with our Table \ref{3dSU6}. The sector with a GNO charge $\left(\frac{1}{2},0,0,0,0, -\frac{1}{2} \right)$ contains two Coulomb branch operators $Y$ and $\tilde{Y}$ which are represented as $\frac{\sqrt{x}}{t^6 u^6}$ and $ \frac{x^{3/4} }{t^6 u^4}$ respectively. The higher order terms can be recognized as the products between the Higgs and Coulomb branch operators.

%%%%%%%%%%%%%%%%%%%%%%%%%%%%%%%%%%%%%%%%%%%%%%%%%%%%%%%%%%%
%%%%%%%%%%%%%%%%%%%%%%%%%%%%%%%%%%%%%%%%%%%%%%%%%%%%%%%%%%%
%%%%%%%%%%%%%%%%%%%%%%%%%%%%%%%%%%%%%%%%%%%%%%%%%%%%%%%%%%%
%%%%%%%%%%%%%%%%%%%%%%%%%%%%%%%%%%%%%%%%%%%%%%%%%%%%%%%%%%%
\section{3d $\mathcal{N}=2$ $USp(6)$ gauge theories}
%%%%%%%%%%%%%%%%%%%%%%%%%%%%%%%%%%%%%%%%%%%%%%%%%%%%%%%%%%%
%%%%%%%%%%%%%%%%%%%%%%%%%%%%%%%%%%%%%%%%%%%%%%%%%%%%%%%%%%%
%%%%%%%%%%%%%%%%%%%%%%%%%%%%%%%%%%%%%%%%%%%%%%%%%%%%%%%%%%%
%%%%%%%%%%%%%%%%%%%%%%%%%%%%%%%%%%%%%%%%%%%%%%%%%%%%%%%%%%%
In four spacetime dimensions, no s-confinement phase is known in the literature for the $USp(2N)$ gauge theories with three-index matters. Those theories flow into non-confining phases along the Higgs branch \cite{Csaki:1996zb}. However, in three spacetime dimensions, we can construct an s-confining theory for a third-oder anti-symmetric tensor in $USp(6)$. Let us consider the 3d $\mathcal{N}=2$ $USp(6)$ gauge theory with three fundamental matters and with one third-order anti-symmetric matter simply denoted as ${\tiny \yng(1,1,1)}$. Table \ref{USptable} shows the matter contents and their quantum numbers. The Higgs branch of the moduli space of vacua is parametrized by 
\begin{gather}
M_{2,0} :=QQ,~~~M_{2,2}:=QA^2Q, \\
B_{3,1} :=Q^3 A,~~~B_{3,3}:=(QA)^3,~~~T_{0,4} :=A^4.
\end{gather}
Table \ref{USptable} also includes the relevant Coulomb brach coordinates which are of importance in our discussion below.

\begin{table}[h]\caption{Quantum numbers of $USp(6)$ with ${\tiny \protect\yng(1,1,1)}$ and $3~ {\tiny \protect\yng(1)}$} 
\begin{center}
\scalebox{0.62}{
  \begin{tabular}{|c||c|c|c|c|c| } \hline
 & $USp(6)$ & $SU(3)$ & $U(1)_Q$ &$U(1)_A$& $U(1)_R$ \\  \hline
$Q$ &${\tiny \yng(1)} $&${\tiny \yng(1)} $&1&0& $R_Q$\\
  $A$&${\tiny \yng(1,1,1)} $&1&0&1&$R_A$  \\[6pt] \hline 
  $M_{2,0}:=QQ$&1&${\tiny \overline{\yng(1)}} $&2&0&$2R_Q$ \\
%$T:=A^2$  &1&1&0&2&$2R_A$ \\
 $B_{3,1}:=Q^3A$ &1&1&3&1&$3R_Q+R_A$ \\ 
$M_{2,2}:=(QA)^2$ &1&${\tiny \yng(2)} $&2&2&$2R_Q+2R_A$ \\
$T_{0,4}:=(A^2)^2$ &1&1&0&4&$4R_A$ \\ 
$B_{3,3}:=(QA)^3$&1&1&3&3&$3R_Q+3R_A$ \\ \hline
%$B_{3,5}:= Q^3A^5$&1&1&3&5&$3R_Q+5R_A$ \\ \hline
$Y_1$&1&1&0&$-2-2\mathrm{sign}(\sigma_1-\sigma_2-2\sigma_3)$&$-2 -2(R_A-1) (1+\mathrm{sign}(\sigma_1-\sigma_2-2\sigma_3))$ \\
$Y_2$&1&1&0&0&$-2$ \\
$Y_3$&1&1&$-3$&$-2+\mathrm{sign}(\sigma_1-\sigma_2-2\sigma_3)$&$-2 -3(R_Q-1)-(R_A-1)(2-\mathrm{sign}(\sigma_1-\sigma_2-2\sigma_3))$ \\
 $Y:=Y_1Y_2Y_3$ $(\sigma_1 >\sigma_2+2\sigma_3)$ &1&1&$-3$&$-4-\mathrm{sign}(\sigma_1-\sigma_2-2\sigma_3)=-5$& $-3R_Q-5R_A+2$ \\
 $\tilde{Y}:=Y_1Y_2^2Y_3^2$ &1&1&$-6$&$-6$& $-6(R_Q+R_A)+2$ \\ \hline
 $\eta :=\Lambda^{b/2}$ &1&1&3&5&$3R_Q+5R_A$ \\ \hline
  \end{tabular}}
  \end{center}\label{USptable}
\end{table}

Let us start by studying the classical Coulomb branch of the moduli space of vacua. Since the $USp(6)$ group has rank three, there are three magnetic monopoles corresponding to the breaking $USp(6) \rightarrow U(1)^3$ at generic points of the Coulomb moduli space. For the monopoles with a simple root $\alpha_i$ $(i=1,2,3)$, we can define the (classical) Coulomb branch operators
\begin{align}
Y_1 & \simeq \exp[\sigma_1-\sigma_2] \\
Y_2 & \simeq\exp[\sigma_2-\sigma_3]  \\
Y_3 & \simeq \exp[2\sigma_3],
\end{align}
where $\sigma_i$ are the diagonal adjoint scalars in a 3d vector multiplet of $USp(6)$, satisfying $\sigma_1 \ge \sigma_2 \ge \sigma_3 \ge 0$ in a certain Weyl chamber. These fields parametrize the classical Coulomb branch which is complex three-dimensional by incorporating the dual photons. Semi-classically, the monopoles can create some non-perturbative superpotential and modify the classical picture. In order to derive the monopole effects, we compute the fermion zero-modes for each monopole. The number of zero-modes is obtained via the Callias index theorem \cite{Callias:1977kg,Weinberg:1979zt,deBoer:1997kr} and the result is summarized in Table \ref{zeromode} below.

\begin{table}[H]\caption{Fermion zero-modes for the $USp(6)$ Coulomb branch} 
\begin{center}
  \begin{tabular}{|c||c|c|c| } \hline
  &adjoint& fundamental & third-order antisymmetric \\ \hline
$Y_1$&2&0&$2+2\mathrm{sign(\sigma_1 -\sigma_2 -2\sigma_3 )}$\\
$Y_2$&2&0&0 \\
$Y_3$&2&1&$2-\mathrm{sign(\sigma_1 -\sigma_2 -2\sigma_3)}$ \\ \hline
$Y:= Y_1 Y_2 Y_3$ $(\sigma_1 > \sigma_2 + 2\sigma_3)$&6&1&5 \\
$\tilde{Y}:=Y_1 Y_2^2 Y_3^2$ &10&2&6 \\ \hline
  \end{tabular}
  \end{center}\label{zeromode}
\end{table}
\noindent From Table \ref{zeromode}, we find that the Coulomb branch should be divided depending on the sign of $\sigma_1 -\sigma_2 -2\sigma_3$. For $\sigma_1 <\sigma_2 + 2\sigma_3$, $Y_1$ and $Y_2$ have two zero-modes only from the gaugino. Hence, the non-perturbative potential $W=\frac{1}{Y_1}+\frac{1}{Y_2}$ is generated and $Y_{1,2}$ are lifted. As a result, the semi-classical moduli space becomes one-dimensional in the region with $\sigma_1 <\sigma_2 + 2\sigma_3$. On the other hand, for $\sigma_1 > \sigma_2 + 2\sigma_3$, $Y_1$ and $Y_2$ have more than two zero-modes. The additional zero-modes come from the matter multiplets. The monopole generates only $W=\frac{1}{Y_2}$. Therefore it is plausible to assume that the Coulomb branch is two-dimensional in the region with $\sigma_1 > \sigma_2 + 2\sigma_3$. From this semi-classical analysis, we introduce two types of operators for the quantum description of the Coulomb moduli
\begin{align}
Y:=Y_1Y_2Y_3,~~~~~~\tilde{Y}:=Y_1Y_2^2 Y_3^2,
\end{align}
where $Y$ is defined for the region of $\sigma_1 > \sigma_2 + 2\sigma_3$ and $\tilde{Y}$ is globally defined in the whole Weyl chamber. It is plausible to use $Y$ coordinate because $Y$ is the globally defined for the 3d $\mathcal{N}=2$ $USp(6)$ theory only with the fundamental matters. $\tilde{Y}$ is also plausible and would be globally defined because this particular combination of the classical coordinates deletes the $\mathrm{sign}(\sigma_1 -\sigma_2 -2\sigma_3)$ dependence. From the zero-mode counting, we can compute the quantum numbers of the Coulomb branch operators as in Table \ref{USptable}. Up to now, the analysis of the Coulomb branch is (semi-)classical. Quantum-mechanically, this picture is still modified. From the quantum numbers of $Y$ and $\tilde{Y}$, we expect that these two coordinates are related in the following way.
\begin{align}
Y \sim \tilde{Y} Q^3 A \sim \tilde{Y} B_{3,1}
\end{align}
This means that $Y$ is a composite operator which consists of $\tilde{Y}$ and $B_{3,1}$. Therefore, we predict that the quantum Coulomb  branch is one-dimensional and described by a globally defined coodinate $\tilde{Y}$. 

By employing the above assumption on the Coulomb branch, we can write down the superpotential consistent with all the symmetries listed in Table \ref{USptable}.
\begin{align}
W = \tilde{Y} \left(M_{2,2}^3 + M_{2,0}^2M_{2,2}T_{0,4} +B^2_{3,1}T_{0,4} +B_{3,3}^2 \right), \label{USP6W}
\end{align}
where we omitted the relative coefficients for simplicity. In order to derive the 4d results, we have to introduce the KK-monopole superpotential $W= \eta Y \sim \eta \tilde{Y} B_{3,1}$. By integrating out the Coulomb branch operators, we find a single quantum-modified constraint.

We can test this dual description in various ways. First, we can easily observe parity anomaly matching between the UV and IR theories. The most non-trivial sector of the parity anomalies is $k^{RR}$ which takes  half odd integers. In order to produce the same anomaly in the dual theory, it is important to introduce only one operator for the Coulomb moduli. This is a weak check of our analysis

As a more non-trivial check, we test a particular Higgs branch direction where the meson $M_{2,0}$ gets an expectation value and the gauge group is broken into $USp(4)$. The low-energy theory becomes a 3d $\mathcal{N}=2$ $USp(4)$ gauge theory with two fundamentals and two anti-symmetric matters. The theory is identical to the 3d $\mathcal{N}=2$ $Spin(5) $ theory with two vectors and two spinors. The matter contents and their quantum numbers are summarized in Table \ref{USp42anti}.
\begin{table}[H]\caption{Quantum numbers of $USp(4)$ with two anti-symmetrics and two fundamentals} 
\begin{center}
\scalebox{1}{
  \begin{tabular}{|c||c|c|c|c|c|c| } \hline
 & $USp(4)$ & $SU(2)_Q$ &$SU(2)_A$& $U(1)_Q$ &$U(1)_A$& $U(1)_R$ \\  \hline
$Q$ &${\tiny \yng(1)} $&${\tiny \yng(1)} $&1&1&0& $R_Q$\\
  $A$&${\tiny \yng(1,1)} $&1&${\tiny \yng(1)} $&0&1&$R_A$  \\[4pt] \hline 
  $M_0:=Q^2$&1&1&1&2&0&$2R_Q$ \\
 $M_1:= QAQ$ &1&1&${\tiny \yng(1)} $&2&1&$2R_Q+R_A$ \\
  $M_2:= QA^2Q$&1&${\tiny \yng(2)} $&1&2&2&$2R_Q+2R_A$ \\
 $T:=A^2$ &1&1&${\tiny \yng(2)} $&0&2&$2R_A$ \\ \hline
  $Y:= V_1V_2$&1&1&1&$-2$&$-4$&$2-2R_Q-4R_A$ \\
  $\tilde{Y}:=V_1V_2^2$&1&1&1&$-4$&$-4$&$2-4R_Q-4R_A$ \\ \hline
  \end{tabular}}
  \end{center}\label{USp42anti}
\end{table}
Let us consider its low-energy dynamics. The classical Coulomb branch of the $USp(4)$ theory is parametrized by
\begin{align}
V_1& \simeq \exp (\sigma_1 -\sigma_2), \\
V_2 & \simeq \exp (\sigma_2)~~~~~~(\sigma_1  \ge \sigma_2 \ge 0)
\end{align}
and there are two monopole configurations correspondingly. The fermion zero-modes around these monopoles are again computed via the Callias index theorem \cite{Callias:1977kg,Weinberg:1979zt,deBoer:1997kr} and summarized in Table \ref{zeromodeUSP4} below.

\begin{table}[H]\caption{Fermion zero-modes for the $USp(4)$ Coulomb branch} 
\begin{center}
  \begin{tabular}{|c||c|c|c| } \hline
  &adjoint& fundamental & antisymmetric \\ \hline
$V_1$&2&0&2\\
$V_2$&2&1&0 \\ \hline
$Y:= V_1V_2$&4&1&2 \\
$\tilde{Y}:=V_1V_2^2$ &6&2&2 \\ \hline
  \end{tabular}
  \end{center}\label{zeromodeUSP4}
\end{table}
The monopoles have more than two fermion zero-modes and do not create any superpotential. We
 expect that the classical two-dimensional Coulomb branch remains flat after including the monopole effects. One might expect that $Y$ and $\tilde{Y}$ describe the Coulomb moduli as in the $USp(6)$ case. However, the symmetry argument again suggests that these two coordinates are related as $Y \sim \tilde{Y}Q^2 \sim \tilde{Y}M_0$. As a result, the quantum Coulomb moduli space is parametrized by a single $\tilde{Y}$ coordinate. The shortening of the Coulomb branch was observed also in the 3d $\mathcal{N}=2$ $Spin(7)$ gauge theory in \cite{Nii:2018tnd}. The effective superpotential becomes
\begin{align}
W=\tilde{Y} \left( M_0^2 T^2  +M_1^2 T  +M_2^2 \right), \label{USp42a}
\end{align}
which is consistent with all the symmetries in Table \ref{USp42anti}. We can reproduce this superpotential from the dual description of the $USp(6)$ theory \eqref{USP6W}. Since the non-abelian global symmetries are modified from $SU(3)$ to $SU(2)_Q \times SU(2)_A$, the composite operators are decomposed as
\begin{gather}
M_{2,2} =: \begin{pmatrix}
(M_2)_{1,1}  & v (M_1)^1 &v(M_1)^2 \\
v (M_1)^1 & v^2 T_{11} &v^2T_{12} \\
v(M_1)^2& v^2 T_{12}& v^2 T_{22}
\end{pmatrix}, \\
B_{3,1} =:v^2 M_0,~~~B_{3,3} =: v^2 (M_2)_{12},~~~T_{0,4} =T+(M_2)_{22},
\end{gather}
where $v^2$ is a vev for $M_{2,0}$. By substituting these expression we find the superpotential \eqref{USp42a} although an additional term $M_0^2 (M_2)_{22}$ is also generated. We expect that this unwanted term vanishes along the RG flow. This is another test of our analysis.

%%%%%%%%%%%%%%%%%%%%%%%%%%%%%%%%%%%%%%%%%%%%%%%%%%%%%%%%%%%
\subsection*{Superconformal Indices for $USp(6)$ with ${\tiny \protect\yng(1,1,1)}$ and $3~{\tiny \protect\yng(1)}$}
%%%%%%%%%%%%%%%%%%%%%%%%%%%%%%%%%%%%%%%%%%%%%%%%%%%%%%%%%%%
Let us study the superconformal indices for the 3d $\mathcal{N}=2$ $USp(6)$ gauge theory with three fundamental matters and with one third-order anti-symmetric tensor. This will be another test of our analysis for the Coulomb branch. Since the dual description has no gauge interaction, the dual index includes only the contributions from the gauge-invariant composite chiral superfields. The dual index is expanded as

\tiny
\begin{align}
I_{dual}&=1+3 t^2 x^{1/4}+\sqrt{x} \left(\frac{1}{t^6 u^6}+6 t^4+t^3 u+6 t^2 u^2+u^4\right)+x^{3/4} \left(10 t^6+3 t^5 u+\frac{3}{t^4 u^6}+18 t^4 u^2+t^3 u^3+3 t^2 u^4\right) \nonumber \\
&\qquad +x \left(\frac{1}{t^{12} u^{12}}+15 t^8+6 t^7 u+37 t^6 u^2+\frac{1}{t^6 u^2}+9 t^5 u^3+27 t^4 u^4+\frac{6}{t^4 u^4}+t^3 u^5+\frac{1}{t^3 u^5}+6 t^2 u^6+\frac{6}{t^2 u^6}+u^8\right) \nonumber \\
&\qquad +x^{5/4} \left(\frac{3}{t^{10} u^{12}}+21 t^{10}+10 t^9 u+63 t^8 u^2+24 t^7 u^3+74 t^6 u^4+9 t^5 u^5+18 t^4 u^6+\frac{3}{t^4 u^2}+t^3 u^7+3 t^2 u^8+\frac{18}{t^2 u^4}+\frac{3}{t u^5}+\frac{10}{u^6}\right) \nonumber \\
&\qquad +x^{3/2} \biggl(\frac{1}{t^{18} u^{18}}+\frac{1}{t^{12} u^8}+28 t^{12}+15 t^{11} u+\frac{6}{t^{10} u^{10}}+96 t^{10} u^2+\frac{1}{t^9 u^{11}}+47 t^9 u^3+\frac{6}{t^8 u^{12}}+150 t^8 u^4+45 t^7 u^5+93 t^6 u^6 \nonumber \\
&\qquad \qquad \qquad \qquad  +\frac{u^2}{t^6}+9 t^5 u^7+27 t^4 u^8+\frac{6}{t^4}+t^3 u^9+\frac{21}{t^2 u^2}+\frac{3 t^2 \left(2 u^{16}+5\right)}{u^6}+\frac{6 t}{u^5}+\frac{6}{t u^3}+\frac{u^{16}+36}{u^4}\biggr) \nonumber \\
&\qquad+x^{7/4} \biggl(\frac{3}{t^{16} u^{18}}+36 t^{14}+21 t^{13} u+136 t^{12} u^2+78 t^{11} u^3+\frac{3}{t^{10} u^8}+255 t^{10} u^4+110 t^9 u^5+\frac{18}{t^8 u^{10}}+237 t^8 u^6+\frac{3}{t^7 u^{11}}+45 t^7 u^7 \nonumber \\
&\qquad \qquad \qquad \qquad  +\frac{10}{t^6 u^{12}}+74 t^6 u^8+9 t^5 u^9+\frac{3 u^2}{t^4}+\frac{3 t^4 \left(6 u^{16}+7\right)}{u^6}+\frac{t^3 \left(u^{16}+10\right)}{u^5}+\frac{3 t^2 \left(u^{16}+20\right)}{u^4}+\frac{15}{t^2}+\frac{18 t}{u^3}+\frac{55}{u^2}\biggr) \nonumber \\
&\qquad +x^2 \biggl(\frac{1}{t^{24} u^{24}}+\frac{1}{t^{18} u^{14}}+\frac{6}{t^{16} u^{16}}+45 t^{16}+\frac{1}{t^{15} u^{17}}+28 t^{15} u+\frac{6}{t^{14} u^{18}}+183 t^{14} u^2+117 t^{13} u^3+390 t^{12} u^4+\frac{1}{t^{12} u^4}+210 t^{11} u^5 \nonumber \\
&\qquad \qquad \qquad +471 t^{10} u^6+\frac{6}{t^{10} u^6}+166 t^9 u^7+276 t^8 u^8+\frac{21}{t^8 u^8}+45 t^7 u^9+\frac{6}{t^7 u^9}+\frac{u^{16}+36}{t^6 u^{10}}+\frac{t^6 \left(93 u^{16}+28\right)}{u^6}+\frac{6}{t^5 u^{11}} \nonumber \\
&\qquad \qquad \qquad +\frac{3 t^5 \left(3 u^{16}+5\right)}{u^5}+\frac{6 u^{16}+15}{t^4 u^{12}}+\frac{9 t^4 \left(3 u^{16}+10\right)}{u^4}+\frac{t^3 \left(u^{16}+36\right)}{u^3}+\frac{21 u^2}{t^2}+\frac{3 t^2 \left(2 u^{16}+35\right)}{u^2}+\frac{15 t}{u}+u^{16}+45\biggr)+\cdots,
\end{align}

\normalsize

\noindent where $t$ and $u$ are the fugacities for the $U(1)_Q \times U(1)_A$ symmetries and we set $R_{Q}=R_A =\frac{1}{8}$ for simplicity. We will reproduce the same index on the electric side below and confirm the validity of the low-energy description \eqref{USP6W}.

The superconformal index on the electric side is decomposed into the indices with different GNO charges. We will list each index below for completeness and give the operator identification for lower terms.

\footnotesize
\begin{align}
I_{electric}^{(0,0,0)} &=1+3 t^2 x^{1/4}+\sqrt{x} \left(6 t^4+t^3 u+6 t^2 u^2+u^4\right)+x^{3/4} \left(10 t^6+3 t^5 u+18 t^4 u^2+t^3 u^3+3 t^2 u^4\right) \nonumber \\
& \qquad +x (15 t^8 + 6 t^7 u + 37 t^6 u^2 + 9 t^5 u^3 + 27 t^4 u^4 + t^3 u^5 + 6 t^2 u^6 + u^8) \nonumber \\
& \qquad +x^{5/4} \left(21 t^{10}+10 t^9 u+63 t^8 u^2+24 t^7 u^3+74 t^6 u^4+9 t^5 u^5+18 t^4 u^6+t^3 u^7+3 t^2 u^8\right) +\cdots  
\end{align}
\normalsize

The index with zero GNO charge contains the Higgs branch operators. The second term $3 t^2 x^{1/4}$ corresponds to a meson $M_{2,0}$. The third term $\sqrt{x} \left(6 t^4+t^3 u+6 t^2 u^2+u^4\right)$ consists of four operators, $M_{2,0}^2, B_{3,1},M_{2,2}$ and $T_{0,4} $, where $M_{2,0}^2$ should be regarded as a symmetric product. $B_{3,3}$ appears as $t^3 u^3 x^{3/4}$ in the fourth term.

\small
\begin{align}
I_{electric}^{\left(\frac{1}{2},0,0 \right)} &=\frac{x}{t^3 u^5}+\frac{3 x^{5/4} (t+u)}{t^2 u^5}+x^{3/2} \left(\frac{6 t}{u^5}+\frac{6}{t u^3}+\frac{9}{u^4}\right)+x^{7/4} \left(\frac{10 t^3}{u^5}+\frac{18 t^2}{u^4}+\frac{18 t}{u^3}+\frac{10}{u^2}\right) \nonumber \\
&\qquad  +x^2 \left(\frac{15 t^5}{u^5}+\frac{30 t^4}{u^4}+\frac{36 t^3}{u^3}+\frac{30 t^2}{u^2}+\frac{15 t}{u}\right)+\cdots 
\end{align}
\normalsize

The index with a GNO charge $\left(\frac{1}{2},0,0 \right)$ starts with the monopole operator $Y$ which is recognized as $\tilde{Y}B_{3,1}$ in our analysis. The second term $\frac{3 x^{5/4} (t+u)}{t^2 u^5}$, at first sight, looks $Y (Q^2 +QA)$. Along the $Y$ direction (or a GNO charge $\left(\frac{1}{2},0,0 \right)$), the gauge group is broken to $USp(4) \times U(1)$. The fundamental and third-order anti-symmetric matters supply the fundamental representations of the unbroken $USp(4)$, which is neutral under the unbroken $U(1)$ symmetry. Therefore, $Q^2$ and $QA$ are regarded as the meson of the $USp(4)$ theory a la \cite{Bashkirov:2011vy}. The coefficient precisely explains the flavor symmetry of $Q$. From the dual theory point of view, these are identified with $\tilde{Y} B_{3,1} M_{2,0}$ and $\tilde{Y} M_{2,0} M_{2,2}$, which is consistent with our analysis for the quantum Coulomb branch $Y \sim \tilde{Y}Q^3A$.

\scriptsize
\begin{align}
I_{electric}^{\left(\frac{1}{2},\frac{1}{2},0 \right)} &=\frac{\sqrt{x}}{t^6 u^6}+\frac{3 x^{3/4}}{t^4 u^6}+x \left(\frac{1}{t^6 u^2}+\frac{6}{t^4 u^4}+\frac{6}{t^2 u^6}\right)+x^{5/4} \left(\frac{3}{t^4 u^2}+\frac{15}{t^2 u^4}+\frac{10}{u^6}\right) +x^{3/2} \left(\frac{u^2}{t^6}+\frac{6}{t^4}+\frac{15 t^2}{u^6}+\frac{21}{t^2 u^2}+\frac{27}{u^4}\right) \nonumber \\
& \qquad +x^{7/4} \left(\frac{21 t^4}{u^6}+\frac{3 u^2}{t^4}+\frac{42 t^2}{u^4}+\frac{15}{t^2}+\frac{45}{u^2}\right)+x^2 \left(\frac{28 t^6}{u^6}+\frac{u^6}{t^6}+\frac{60 t^4}{u^4}+\frac{6 u^4}{t^4}+\frac{75 t^2}{u^2}+\frac{21 u^2}{t^2}+55\right)+ \cdots 
\end{align}
\normalsize

The index with a GNO charge $\left(\frac{1}{2},\frac{1}{2},0 \right)$ starts with the monopole operator $\tilde{Y}$ which is represented as $\frac{\sqrt{x}}{t^6 u^6}$. The second term $\frac{3 x^{3/4}}{t^4 u^6}$ corresponds to $\tilde{Y}M_{2,0}$ and the third term $x \left(\frac{1}{t^6 u^2}+\frac{6}{t^4 u^4}+\frac{6}{t^2 u^6}\right)$ comes from $\tilde{Y} (B_{0,4}+B_{2,2}+M_{2,0}^2) $.
Up to $O(x^2)$, the following sectors should be summed up and we observe exact matching between the electric and magnetic indices.

\small
\begin{align}
I_{electric}^{(1,0,0)}&= \frac{x^2}{t^6 u^{10}}+\cdots, \\
I_{electric}^{(1,1/2,0)}&=\frac{x^{3/2}}{t^9 u^{11}}+\frac{3 x^{7/4} (t+u)}{t^8 u^{11}}+\frac{2 x^2 \left(3 t^2+4 t u+3 u^2\right)}{t^7 u^{11}}+\cdots, \\
I_{electric}^{(1,1,0)}&=\frac{x}{t^{12} u^{12}}+\frac{3 x^{5/4}}{t^{10} u^{12}}+x^{3/2} \left(\frac{1}{t^{12} u^8}+\frac{6}{t^{10} u^{10}}+\frac{6}{t^8 u^{12}}\right)+x^{7/4} \left(\frac{3}{t^{10} u^8}+\frac{15}{t^8 u^{10}}+\frac{10}{t^6 u^{12}}\right) \nonumber \\
&\qquad +x^2 \left(\frac{1}{t^{12} u^4}+\frac{6}{t^{10} u^6}+\frac{21}{t^8 u^8}+\frac{27}{t^6 u^{10}}+\frac{15}{t^4 u^{12}}\right)+\cdots, \\
I_{electric}^{(3/2,1,0)}&=\frac{x^2}{t^{15} u^{17}} +\cdots, \\
I_{electric}^{(3/2,3/2,0)}&=\frac{x^{3/2}}{t^{18} u^{18}}+\frac{3 x^{7/4}}{t^{16} u^{18}}+x^2 \left(\frac{1}{t^{18} u^{14}}+\frac{6}{t^{16} u^{16}}+\frac{6}{t^{14} u^{18}}\right)+\cdots, \\
I_{electric}^{(2,2,0)}&=\frac{x^2}{t^{24} u^{24}}+\cdots.
\end{align}
\normalsize

%%%%%%%%%%%%%%%%%%%%%%%%%%%%%%%%%%%%%%%%%%%%%%%%%%%%%%%%%%%
\subsection*{Superconformal Indices for $USp(4)$ with $2~{\tiny \protect\yng(1,1)}$ and $2~{\tiny \protect\yng(1)}$}
%%%%%%%%%%%%%%%%%%%%%%%%%%%%%%%%%%%%%%%%%%%%%%%%%%%%%%%%%%%
Finally, we also compute the superconformal indices for the 3d $\mathcal{N}=2$ $USp(4)$ gauge theory with two antisymmetric tensors and two fundamentals. Since the theory appears from the Higgs branch of the $USp(6)$ with ${\tiny \yng(1,1,1)}$ and $3~{\tiny \yng(1)}$, it is expected that the $USp(4)$ theory also shows the s-confinement as we derived the exact superpotential \eqref{USp42a}. It is worth investigating the index and understanding the low-lying operators in the chiral ring. It is also valuable to confirm the euqivalence of the indices between the $USp(4)$ theory and the magnetic confined description. We start with the SCI of the dual description.    

\tiny
\begin{align}
I_{dual} &=1+x^{1/4} \left(t^2+3 u^2\right)+2 t^2 u x^{3/8}+\sqrt{x} \left(t^4+6 t^2 u^2+6 u^4\right)+2 t^2 u x^{5/8} \left(t^2+3 u^2\right)+x^{3/4} \left(t^6+9 t^4 u^2+15 t^2 u^4+10 u^6\right) \nonumber \\
&+2 t^2 u x^{7/8} \left(t^4+6 t^2 u^2+6 u^4\right)+x \left(t^8+9 t^6 u^2+29 t^4 u^4+\frac{1}{t^4 u^4}+28 t^2 u^6+15 u^8\right)+2 t^2 u x^{9/8} \left(t^6+8 t^4 u^2+15 t^2 u^4+10 u^6\right) \nonumber \\
&+x^{5/4} \left(t^{10}+9 t^8 u^2+38 t^6 u^4+61 t^4 u^6+\frac{3}{t^4 u^2}+45 t^2 u^8+\frac{1}{t^2 u^4}+21 u^{10}\right)+x^{11/8} \left(2 t^{10} u+16 t^8 u^3+52 t^6 u^5+56 t^4 u^7+30 t^2 u^9+\frac{2}{t^2 u^3}\right)  \nonumber \\
&+x^{3/2} \left(t^{12}+9 t^{10} u^2+43 t^8 u^4+95 t^6 u^6+105 t^4 u^8+\frac{6}{t^4}+66 t^2 u^{10}+\frac{3}{t^2 u^2}+28 u^{12}+\frac{1}{u^4}\right)  \nonumber \\
&+x^{13/8} \left(2 t^{12} u+16 t^{10} u^3+64 t^8 u^5+110 t^6 u^7+90 t^4 u^9+42 t^2 u^{11}+\frac{4}{t^2 u}+\frac{2}{u^3}\right)  \nonumber \\
&+x^{7/4} \left(t^{14}+9 t^{12} u^2+43 t^{10} u^4+125 t^8 u^6+180 t^6 u^8+161 t^4 u^{10}+\frac{10 u^2}{t^4}+t^2 \left(91 u^{12}+\frac{1}{u^4}\right)+\frac{5}{t^2}+\frac{36 u^{16}+3}{u^2}\right) \nonumber \\
&+x^{15/8} \left(2 t^{14} u+16 t^{12} u^3+70 t^{10} u^5+160 t^8 u^7+190 t^6 u^9+132 t^4 u^{11}+\frac{2 t^2 \left(28 u^{16}+1\right)}{u^3}+\frac{6 u}{t^2}+\frac{4}{u}\right) \nonumber \\
&+x^2 \left(t^{16}+9 t^{14} u^2+43 t^{12} u^4+140 t^{10} u^6+264 t^8 u^8+\frac{1}{t^8 u^8}+293 t^6 u^{10}+\frac{15 u^4}{t^4}+t^4 \left(229 u^{12}+\frac{1}{u^4}\right)+\frac{7 u^2}{t^2}+\frac{3 t^2 \left(40 u^{16}+1\right)}{u^2}+45 u^{16}-3\right)+\cdots,
\end{align}
\normalsize

\noindent where we introduced the fugacities $(t,u)$ for the $U(1)_Q \times U(1)_A$ global abelian symmetry and set $R_Q=R_A=\frac{1}{8}$ for simplicity.  
The dual theory has no gauge interaction and only the chiral superfields contribute to the index. The Higgs branch operators $M_0,M_1,M_2$ and $T$ are represented as $t^2x^{1/4},2t^2 u x^{3/8},3t^2u^2 x^{1/2}$ and $3 u^2 x^{1/4}$ in the index above. The Coulomb branch operator $\tilde{Y}$ is denoted as $\frac{x}{t^4 u^4}$. The higher order terms are recognized as the symmetric products of these fields with constraints from the superpotential \eqref{USp42a}.

Next, we consider the index on the electric side. Since the electric (UV) description contains the gauge interaction of $USp(4)$, the index is decomposed into the indices with different GNO charges. For completeness, we will list each index separately. Up to $O(x^2)$, we have to sum up the following sectors and observe a complete agreement with the magnetic side.

\scriptsize
\begin{align}
I_{electric}^{(0,0)} &= 1+x^{1/4} \left(t^2+3 u^2\right)+2 t^2 u x^{3/8}+\sqrt{x} \left(t^4+6 t^2 u^2+6 u^4\right)+x^{5/8} \left(2 t^4 u+6 t^2 u^3\right)+x^{3/4} \left(t^6+9 t^4 u^2+15 t^2 u^4+10 u^6\right)\nonumber \\
&+x^{7/8} \left(2 t^6 u+12 t^4 u^3+12 t^2 u^5\right)+x \left(t^8+9 t^6 u^2+29 t^4 u^4+28 t^2 u^6+15 u^8\right)+x^{9/8} \left(2 t^8 u+16 t^6 u^3+30 t^4 u^5+20 t^2 u^7\right) \nonumber \\
&+x^{5/4} \left(t^{10}+9 t^8 u^2+38 t^6 u^4+61 t^4 u^6+45 t^2 u^8+21 u^{10}\right)+x^{11/8} \left(2 t^{10} u+16 t^8 u^3+52 t^6 u^5+56 t^4 u^7+30 t^2 u^9\right) \nonumber \\
&+x^{3/2} \left(t^{12}+9 t^{10} u^2+43 t^8 u^4+95 t^6 u^6+105 t^4 u^8+66 t^2 u^{10}+28 u^{12}\right)\nonumber \\
&+x^{13/8} \left(2 t^{12} u+16 t^{10} u^3+64 t^8 u^5+110 t^6 u^7+90 t^4 u^9+42 t^2 u^{11}\right) \nonumber \\
&+x^{7/4} \left(t^{14}+9 t^{12} u^2+43 t^{10} u^4+125 t^8 u^6+180 t^6 u^8+161 t^4 u^{10}+91 t^2 u^{12}+36 u^{14}\right)  \nonumber \\
&+x^{15/8} \left(2 t^{14} u+16 t^{12} u^3+70 t^{10} u^5+160 t^8 u^7+190 t^6 u^9+132 t^4 u^{11}+56 t^2 u^{13}\right) \nonumber \\
&+x^2 \left(t^{16}+9 t^{14} u^2+43 t^{12} u^4+140 t^{10} u^6+264 t^8 u^8+293 t^6 u^{10}+229 t^4 u^{12}+120 t^2 u^{14}+45 u^{16}-8\right)+\cdots 
\end{align}
\small
\begin{align}
I_{electric}^{\left(\frac{1}{2},0  \right)} &= \frac{x^{5/4}}{t^2 u^4}+\frac{2 x^{11/8}}{t^2 u^3}+x^{3/2} \left(\frac{3}{t^2 u^2}+\frac{1}{u^4}\right)+x^{13/8} \left(\frac{4}{t^2 u}+\frac{2}{u^3}\right)+x^{7/4} \left(\frac{t^2}{u^4}+\frac{5}{t^2}+\frac{3}{u^2}\right) \nonumber \\
&\qquad \qquad +x^{15/8} \left(\frac{2 t^2}{u^3}+\frac{6 u}{t^2}+\frac{4}{u}\right)+x^2 \left(\frac{t^4}{u^4}+\frac{3 t^2}{u^2}+\frac{7 u^2}{t^2}+5\right)+\cdots \\
I_{electric}^{\left(\frac{1}{2},  \frac{1}{2} \right)} &=\frac{x}{t^4 u^4}+\frac{3 x^{5/4}}{t^4 u^2}+\frac{6 x^{3/2}}{t^4}+\frac{10 u^2 x^{7/4}}{t^4}+\frac{15 u^4 x^2}{t^4}+\cdots \\
I_{electric}^{(1,1)} &=\frac{x^2}{t^8 u^8}+\cdots
\end{align}
\normalsize

\noindent The index with zero GNO charge contains only the Higgs branch coordinates and their symmetric products. The index with a GNO charge $\left(\frac{1}{2},0  \right)$ is classically regarded as the Coulomb branch $Y$ but it is identified with $\tilde{Y}M_0$. The second term $\frac{2 x^{11/8}}{t^2 u^3}$ corresponds to $Y \times 2ux^{1/8} \sim \tilde{Y} Q^2A \sim \tilde{Y}M_1$. This is consistent with our analysis which claims $Y \sim \tilde{Y}Q^2$. 
The index with a GNO charge $\left(\frac{1}{2}, \frac{1}{2}  \right)$ contains the Coulomb branch operator $\tilde{Y}$. The first term $\frac{x}{t^4 u^4}$ precisely exhibits the quantum numbers of $\tilde{Y}$. The proceeding terms are identified with $\tilde{Y} T^n$, where $T^n$ is a symmetric product of $T$. By summing up all the sectors above, we reproduce the magnetic superconformal index. This again confirms the validity of our study.

%\newpage
%%%%%%%%%%%%%%%%%%%%%%%%%%%%%%%%%%%%%%%%%%%%%%%%%%%%%%%%%%
%%%%%%%%%%%%%%%%%%%%%%%%%%%%%%%%%%%%%%%%%%%%%%%%%%%%%%%%%%
\section{Summary and Discussion}
%%%%%%%%%%%%%%%%%%%%%%%%%%%%%%%%%%%%%%%%%%%%%%%%%%%%%%%%%%
%%%%%%%%%%%%%%%%%%%%%%%%%%%%%%%%%%%%%%%%%%%%%%%%%%%%%%%%%%
%[In this work][CBの次元について言及せよ]
In this paper, we investigated the low-energy dynamics for the 3d $\mathcal{N}=2$ $SU(6)$ and $USp(6)$ gauge theories with a three-index matter by paying a special attention to the s-confinement phases. For the $SU(6)$ case, we found an s-confining description for the theory with ${\tiny \yng(1,1,1)}$ and $3~({\tiny \yng(1)} +{\tiny \overline{\yng(1)} })$ and derived the exact superpotential which governs the confined degrees of freedom. The quantum Coulomb branch is complex two-dimensional and described by $Y$ and $\tilde{Y}$. The 3d s-confinement for the $SU(6)$ theory was independently derived from the corresponding 4d s-confinement and also beautifully connected to the 4d quantum-deformed moduli space via the KK-monopole. As consistency checks, we studied the low-energy limit along the Higgs branch and computed the superconformal indices. 

For the $USp(6)$ case, the 3d $\mathcal{N}=2$ $USp(6)$ gauge theory with ${\tiny \yng(1,1,1)}$ and $3~{\tiny \yng(1)}$ showed the 3d s-confinement while the s-confinement does not occur for the corresponding 4d theory. Although the classical analysis suggests two Coulomb branch operators, the quantum Coulomb branch of the $USp(6)$ theory is described by a single operator $\tilde{Y}$ which is globally defined. We tested the $USp(6)$ s-confinement by flowing to the Higgs branch and  computing the superconformal indices. As a by-product, we found that the 3d $\mathcal{N}=2$ $USp(4)$ theory with $2~{\tiny \yng(1,1)}$ and $2~{\tiny \yng(1)}$ is s-cofining, which has not been known in the literature.

%[What we did not do in this paper]
Our analysis assumed the correct coordinates of the Coulomb moduli from various consistencies. For the $SU(6)$ case, the connection between the 4d and 3d theories strongly supports our prediction of the (quantum) Coulomb moduli. For the $USp(6)$ case, the parity anomaly matching weakly suggested that there is a one-dimensional Coulomb branch un-lifted. The SCI also supported these assumptions. It is quite preferable to gain better understanding and more rigorous analysis of the Coulomb branch. This will be a future direction of our study.

%[other s-con, other matters]
Although we here found the two s-confinement phases including three-index matters, it is not exhausting all possibilities for the s-confinement of three-index matters. Furthermore, it is still unclear how to more systematically understand the low-energy dynamics for the theory with multi-index (more than three indices) matters. It is quite interesting to search for more and more confining phases in 3d SUSY gauge theories. The (semi-)classical analysis of the Coulomb branch and the SCI calculation would help us to understand it.

%[Seiberg duality]
We restricted our attention to the s-confinement phases for three-index anti-symmetric matters. That is why the number of (anti-)fundamental quarks are restricted to the particular values. 
In both cases, the number of fundamental representations was three. It is straightforward to obtain the dynamics for the lower number of fundamentals by integrating out the quarks via the complex mass deformation. On the other hand, for larger number of fundamentals, we expect that a certain Seiberg duality gives the correct low-energy dynamics as in the 4d case \cite{Csaki:1997cu}.  It is quite tempting to explore the 3d Seiberg duality for multi-index matters. It is also interesting to study the theory with multiple three-index matters or more general multi-index matters.

%[Future direction]

%%%%%%%%%%%%%%%%%%%%%%%%%%%%%%%%%%%%%%%%%%%%%%%%%%%%%%%%%%
%%%%%%%%%%%%%%%%%%%%%%%%%%%%%%%%%%%%%%%%%%%%%%%%%%%%%%%%%%
\section*{Acknowledgments}
%%%%%%%%%%%%%%%%%%%%%%%%%%%%%%%%%%%%%%%%%%%%%%%%%%%%%%%%%%
%%%%%%%%%%%%%%%%%%%%%%%%%%%%%%%%%%%%%%%%%%%%%%%%%%%%%%%%%%
This work is supported by the Swiss National Science Foundation (SNF) under grant number PP00P2\_157571/1.

\bibliographystyle{ieeetr}
\bibliography{3index}

\end{document}